\documentclass[onecolumn,showpacs,preprintnumbers]{revtex4}
\usepackage{amssymb}

\usepackage{amsfonts}
\usepackage{amsmath}
\usepackage{graphicx}

\input{tcilatex}
\begin{document}

\date{\today }
\title{EPR measurement and the origin of cosmic density fluctuations}
\author{Masahiro Morikawa \thanks{%
hiro@phys.ocha.ac.jp}}
\affiliation{Department of Physics, Ochanomizu University, 2-1-1 Ohtuka, Bunkyo, Tokyo
112-8610, Japan}

\begin{abstract}
We explore consistent application of quantum mechanics to the objects in the
Universe and in laboratories. The measurement dynamics in quantum mechanics
is modeled as a physical process of spontaneous symmetry breaking (SSB)
which is described by the generalized effective action method. A violation
of the Bell inequality is observed in this model and the generation of the
density fluctuations in the early Universe is described as the SSB process
of the spatially translational symmetry. 
\end{abstract}

\maketitle

\section{Introduction}

Quantum mechanics is an excellent theory to describe the Universe. For
example, the mass and the radius of typical planets are given by%
\begin{equation}
M=\frac{e^{3}k_{0}^{3/2}}{2\sqrt{2}G^{3/2}m_{p}^{2}}\approx 8.1\times 10^{26}%
\text{Kg},\;\;\;R=\frac{\hbar ^{2}}{3\sqrt{2Gk_{0}}em_{e}m_{p}}\approx
1.0\times 10^{7}\text{m}
\end{equation}%
where $m_{e}$ and $m_{p}$ are the electron and proton masses, respectively.
These expressions are derived by minimizing the energy $E=p^{2}/\left(
2m_{e}\right) -k_{0}e^{2}/x$ of an electron in the Hydrogen atom using the
uncertainty principle $\left( \Delta x\right) \left( \Delta p\right) \approx
\hbar /2.$ Moreover, the scale of typical stars, which are in equilibrium
with the nuclear fusion process, is given by the balance of the energy
uncertainty $E\approx p^{2}/\left( 2m_{p}\right) $ and the Coulomb barrier $%
k_{0}e^{2}/x$. This balance yields the approximate temperature $T\approx
5\times 10^{8}$K. Further balance of the gravitational energy and the
Coulomb energy in the whole star yields the expression 
\begin{equation}
M=\frac{e^{3}k_{0}^{3/2}}{8G^{3/2}m_{e}^{3/2}m_{p}^{1/2}}\approx 2.3\times
10^{31}\text{Kg},\;\;\;R=\frac{39.5\sqrt{eV}\hbar ^{3}}{e^{3}\sqrt{G}%
k_{0}^{3/2}m_{e}^{3/2}m_{p}}\approx 3.2\times 10^{8}\text{m}
\end{equation}%
for a typical star. The appearance of $\hbar $ in the above radii $R$
becomes more evident in the degenerate stars, in which both the mass and the
radius are proportional to $\hbar ^{3/2}.$ These observations indicate the
fact that quantum mechanics is indispensable in the Universe. There is yet
another implication in the case of bosons \cite{BEC}. If some kind of light
boson field ($m\leq 20$eV) dominates the Universe as dark matter, then it
can yield Bose-Einstein condensation (BEC) which behaves as a classical
scalar field $\varphi \left( x\right) $. This condensed field successfully
describes the dynamics of dark energy \cite{BEC}. Quantum mechanics is thus
fully compatible with astrophysics in the Universe.

However, if we consider the operational description of quantum mechanics,
then the whole argument becomes quite complicated. We cannot describe the
collapse of the wave function of the Universe since no explicit observer is
there. The whole structure in the Universe might fall into big illusion.

More serious problem arises when we describe the origin of the density
fluctuations from quantum mechanics. In the inflationary era in the early
Universe, the k-mode of the light scalar field behaves as $v_{k}^{\prime
\prime }+\left( k^{2}-2H^{2}e^{2Ht}\right) v_{k}=0$, whose solution in de
Sitter invariant vacuum behaves as $v_{k}\approx k^{-1/2}\left(
1+ik^{-1}He^{Ht}\right) .$ A standard calculation for the power spectrum $%
P\left( k\right) $ implicitly assumes the identification of the quantum
correlation of the scalar field $\left\langle \hat{\phi}\left( x\right) \hat{%
\phi}\left( y\right) \right\rangle _{k}$ and the statistical fluctuation $%
\left\vert \delta \varphi \left( k\right) \right\vert ^{2}$ \cite{pad1993} 
\cite{weinberg2008}.The latter violates the spatial translational invariance
while the former does not. This symmetry breaking of the translational
invariance is indispensable for the generation of classical density
fluctuations \cite{morikawa1987} \cite{leon2011}.

It is important to notice that the origin of the problem is common with the
laboratory quantum mechanics. The problem stems from the lack of physical
description of the quantum measurement apparatus, which should have
autonomous dynamics independent from an artificial operator. Therefore we
first consider a physical dynamics of the quantum measurement process and
then apply it to the laboratory measurement and to the early Universe.

\section{Physics of quantum measurement}

Our proposal is that the measurement is a spontaneous symmetry breaking
(SSB) process and is described by the generalized effective action method 
\cite{morikawa1986} \cite{morikawa1995}. The classical order parameter of
the field which cause SSB corresponds to the meter of the measurement
apparatus. The generalized effective action fully describes the dynamics of
this field and the order parameter. This field couples to the system and
affects the evolution of the density matrix of the system. The finite filed $%
\varphi =\left\langle \hat{\phi}\right\rangle \neq 0$ means the state of the
field $\hat{\phi}$ is a coherent state or BEC state. The macroscopic
condensation guarantees the classical property of the field $\varphi .$

A key observation how the generalized effective action describes the SSB is
to consider the entropy $S_{e}=-k_{B}$Tr$\left( \rho \ln \rho \right) ,$ and
the action $S_{a}=-i\hbar \ln \Phi ,$ where $\rho =\left\vert \Phi
\right\rangle \left\langle \Phi \right\vert .$ Then the entropy becomes%
\begin{equation}
S_{e}=\frac{k_{B}}{\hbar }\func{Im}\left[ \sum \left( S_{a}-S_{a}^{\ast (%
\text{time reversal})}\right) \right] ,
\end{equation}%
where the quantity $\sum \left( S_{a}-S_{a}^{\ast (\text{time reversal}%
)}\right) $ in the right hand side is almost the generalized effective
action $\tilde{\Gamma}$. This observation suggests that $\func{Im}\tilde{%
\Gamma}$ describes statistical property of the system while $\func{Re}\tilde{%
\Gamma}$ describes the ordinary dynamics.

There is a neat formalism of the generalized effective action method \cite%
{morikawa1986} \cite{morikawa1995}. We simply review the essence of it in
the case of real scale field. The effective action $\tilde{\Gamma}$ is a
functional of the classical field $\varphi _{c}\equiv \left( \varphi
_{+}+\varphi _{-}\right) /2$ and $\varphi _{\Delta }\equiv \varphi
_{+}-\varphi _{-}:$ $\tilde{\Gamma}\left[ \varphi _{c},\varphi _{\Delta }%
\right] $, where the support of $\varphi _{+}$ is the ordinary time contour $%
-\infty \rightarrow \infty $ while that of $\varphi _{-}$ is the reversed
time contour $\infty \rightarrow -\infty $. The imaginary part is generally
even in $\varphi _{\Delta }$ and within the Gaussian approximation, 
\begin{equation}
\func{Im}\tilde{\Gamma}=\frac{1}{2}\int \int \varphi _{\Delta }\left(
x\right) B\left( x-y\right) \varphi _{\Delta }\left( y\right) +...
\end{equation}%
where the kernel $B\left( x-y\right) $ is positive definite. Therefore by
introducing real auxiliary field $\xi \left( x\right) ,$ we have 
\begin{equation}
\exp \left( i\tilde{\Gamma}\right) =\int \left[ d\xi \right] P\left[ \xi %
\right] \exp \left[ i\func{Re}\tilde{\Gamma}+i\int \xi \varphi _{\Delta }%
\right]
\end{equation}%
where $P\left[ \xi \right] \equiv \exp \left[ -\frac{1}{2}\int \int \xi
B^{-1}\xi \right] $ can be interpreted as statistical weight function$.$ On
the other hand $\func{Re}\tilde{\Gamma}+\xi \varphi _{\Delta }\equiv S_{%
\text{eff}}$ is real and we can apply the least action principle as usual $%
\delta S_{\text{eff}}/\delta \varphi _{\Delta }\left( x\right) |_{\varphi
_{\Delta }=0}=-J_{c}$. This yields a probabilistic differential equation,
which is real and causal: 
\begin{equation}
\left( \Box +m^{2}\right) \varphi _{c}\left( x\right) =-V^{\prime
}+\int_{-\infty }^{t}dt^{\prime }\int d^{3}x^{\prime }A\left( x-x^{\prime
}\right) \varphi _{c}\left( x^{\prime }\right) +\xi \left( x\right) ,
\end{equation}%
where $A\left( x-x^{\prime }\right) $ represents retarded kernel. The
stochastic force term $\xi \left( x\right) $, which arose from $\func{Im}%
\tilde{\Gamma}$, triggers SSB and the classical field $\varphi _{c}\left(
x\right) $ eventually controls the measurement process as we will see soon.

The most simple model of spin ($1/2$) measurement by the field $\phi $ \cite%
{morikawa2006} will be given by the Lagrangian. 
\begin{equation}
L=\frac{1}{2}\left( \partial _{\mu }\phi \right) ^{2}-\frac{m^{2}}{2}\phi
^{2}-\frac{\lambda }{4!}\phi ^{4}+\mu \phi \vec{S}\cdot \vec{B}+\left( \text{%
bath}\right) ,  \label{L}
\end{equation}%
where $m^{2}<0,$ $\lambda >0,$ $\mu >0.$ From the method of generalized
effective action, we have a set of equations, for the order parameter or the
meter of the apparatus, 
\begin{equation}
\frac{d\varphi }{dt}=\gamma \varphi -\frac{\lambda }{3!}\varphi ^{3}+\mu
\left\langle \vec{S}\right\rangle \cdot \vec{B}+\xi ,  \label{phidot}
\end{equation}%
where we set $\left\langle \xi \left( t\right) \xi \left( t^{\prime }\right)
\right\rangle =\varepsilon \delta \left( t-t^{\prime }\right) $ for
simplicity, and for the density matrix of the system,%
\begin{equation}
\frac{d\rho }{dt}=-i\omega \left[ S_{3},\rho \right] +a\left[ S_{+}\rho
,S_{-}\right] +b\left[ S_{-}\rho ,S_{+}\right] +c\left[ S_{3}\rho ,S_{3}%
\right] +h.c.  \label{eqrho}
\end{equation}%
where $a,b,$ and $c$ are the bath-correlation functions including the order
parameter $\varphi ,$ and $a=$ $\exp \left( -\hbar \mu \varphi \left(
t\right) B/\left( kT\right) \right) b.$ If initially $\left\langle \vec{S}%
\right\rangle \cdot \vec{B}>0$, then the linear bias forces $\varphi $ to
move toward $\varphi _{+}>0.$ Then the effective temperature reduces $%
kT/\left( \mu \varphi \left( t\right) \right) \rightarrow 0$, and the spin
falls into the corresponding ground state and becomes pure up-state. This
further makes initial $\left\langle \vec{S}\right\rangle \cdot \vec{B}$
increase. If initially $\left\langle \vec{S}\right\rangle \cdot \vec{B}>0$,
then the opposite situation makes the spin purely down-state with $\varphi
\rightarrow \varphi _{-}<0.$ This positive feedback is the essence of the
quantum measurement. Eq.(\ref{phidot}) describes the process of SSB
violating the original symmetry $\phi \rightarrow -\phi $ in the Lagrangian
Eq.(\ref{L}). This method predicts the finite time scale of the measurement $%
t_{0}=\left( 2\gamma \right) ^{-1}\ln \left[ \left( g/\gamma \right) \left(
\delta ^{2}+\left( \varepsilon /\gamma \right) \right) \right] ^{-1}$, where 
$\delta \equiv \left( \mu /\gamma \right) \left\langle \vec{S}\right\rangle
\cdot \vec{B}$.

\section{Quantum mechanics in laboratories}

We now apply this model to the EPR measurement of two spins $\vec{S}_{1},%
\vec{S}_{2}$ \cite{einstein1935}. We set spatially separated local
measurement apparatus composed from the fields $\phi _{1},\phi _{2}$ for
spins $\vec{S}_{1},\vec{S}_{2}$ respectively, with the same Lagrangian for
each set. We are especially interested in the maximally entangled initial
state of spins $\left\vert \psi \right\rangle =\left( \left\vert \uparrow
\right\rangle _{1}\otimes \left\vert \downarrow \right\rangle
_{2}-\left\vert \downarrow \right\rangle _{1}\otimes \left\vert \uparrow
\right\rangle _{2}\right) /\sqrt{2}$. We have the evolution equation of the
order parameters or the meters of the measurement apparatus $\varphi _{1}$
and $\varphi _{2}$ :%
\begin{eqnarray}
\dot{\varphi}_{1} &=&\gamma \varphi _{1}-\frac{\lambda }{3!}\varphi
_{1}^{3}+\mu \mbox{Tr}\left[ \rho \vec{S}_{1}\cdot \vec{B}_{1}\right] +\vec{%
\xi}_{1}\cdot \vec{B}_{1}+\vec{\xi}_{1}^{\left( 0\right) },  \label{eqphi} \\
\dot{\varphi}_{2} &=&\gamma \varphi _{2}-\frac{\lambda }{3!}\varphi
_{2}^{3}+\mu \mbox{Tr}\left[ \rho \vec{S}_{2}\cdot \vec{B}_{2}\right] +\vec{%
\xi}_{2}\cdot \vec{B}_{2}+\vec{\xi}_{2}^{\left( 0\right) },  \notag
\end{eqnarray}%
where the imaginary part of the effective action yields the statistical
correlation for the classical stochastic forces: $\left\langle {\xi
_{1}{}_{i}\xi _{2}{}_{j}}\right\rangle =Tr\left( \rho \left( t\right)
S_{i}^{(1)}S_{j}^{(2)}\right) $. The fields $\vec{\xi}_{1}^{\left( 0\right)
} $ and $\vec{\xi}_{2}^{\left( 0\right) }$ are the previous stochastic
forces arising from the bath for each apparatus. If the initial state is the
above most entangle state $\left\vert \psi \right\rangle $, then we have the
complete anti-correlation:$\left\langle {\xi _{1}{}_{i}\xi _{2}{}_{j}}%
\right\rangle =-\delta _{ij}.$ For the system state $\rho \left( t\right)
\equiv \rho _{1}\otimes \rho _{2}$, the same equation (\ref{eqrho}) holds.
This set of equations (\ref{eqrho}, \ref{eqphi}) determines the full quantum
mechanical time evolution including the measurement process as previously.

We can observe a violation of Bell-inequality in this formalism. The
probability that the detector apparatus-1 reads the spin-up state under the
applied magnetic field $\vec{B}_{1}$ becomes \cite{morikawa2006} 
\begin{equation}
P_{+}^{\left( 1\right) }\left( \vec{B}_{1}\right) =\int_{0}^{\infty }P\left(
\varphi _{1},t\right) d\varphi =\frac{1+\func{erf}\left( \delta /\sqrt{%
2\varepsilon \left( t\right) }\right) }{2}
\end{equation}%
where $P\left( \varphi _{1},t\right) $ is the probability distribution
function of the order parameter $\varphi _{1}$ at time t. The expression in
the most right hand side is given by the approximate solution of the
Fokker-Planck equation corresponding to (\ref{eqphi}). In the above, $\delta
=\vec{\xi}_{1}\cdot \vec{B}_{1}/\gamma ,$ $\sqrt{2\varepsilon \left(
t\right) }\approx \gamma ^{-1/2}.$ We also have the similar expression for $%
P_{-}^{\left( 1\right) }\left( \vec{B}_{1}\right) $ and $P_{\pm }^{\left(
2\right) }\left( \vec{B}_{2}\right) .$ Defining 
\begin{equation}
P_{++}\equiv \left\langle P_{+}^{\left( 1\right) }\left( \vec{B}_{1}\right)
P_{+}^{\left( 2\right) }\left( \vec{B}_{2}\right) \right\rangle
,\;P_{+-}\equiv \left\langle P_{+}^{\left( 1\right) }\left( \vec{B}%
_{1}\right) P_{-}^{\left( 2\right) }\left( \vec{B}_{2}\right) \right\rangle
,...
\end{equation}
and arranging the magnetic field strengths as $\left\vert \vec{B}%
_{1}\right\vert \left\vert \vec{B}_{2}\right\vert =\gamma $, we have 
\begin{eqnarray}
C\left( \vec{B}_{1},\vec{B}_{2}\right) &=&P_{++}+P_{--}-\left(
P_{+-}+P_{-+}\right) \\
&=&\left\langle \func{erf}\left( \gamma ^{-1/2}\vec{\xi}_{1}\cdot \vec{B}%
_{1}\right) \func{erf}\left( \gamma ^{-1/2}\vec{\xi}_{2}\cdot \vec{B}%
_{2}\right) \right\rangle  \notag \\
&\approx &-\gamma ^{-1}\vec{B}_{1}\cdot \vec{B}_{2}=-\cos \theta _{12} 
\notag
\end{eqnarray}%
where $\theta _{12}$ is the angle between the directions of $\vec{B}_{1}$
and $\vec{B}_{2}$. For some range of the configuration of $\vec{B}_{1},\vec{B%
}_{2},\vec{B}_{2}^{\prime }$ and $\vec{B}_{2}^{\prime }$, we have the
relation%
\begin{equation}
\left\vert C\left( \vec{B}_{1},\vec{B}_{2}\right) +C\left( \vec{B}%
_{1}^{\prime },\vec{B}_{2}\right) +C\left( \vec{B}_{1},\vec{B}_{2}^{\prime
}\right) -C\left( \vec{B}_{1}^{\prime },\vec{B}_{2}^{\prime }\right)
\right\vert >2,
\end{equation}%
which shows a violation of Bell-inequality. Thus the present formalism does
not belong to the category of hidden variable theories, but being just
another description of quantum mechanics.

\section{Quantum mechanics in the Universe}

In the case of early Universe, the generation process of the primordial
density fluctuations is very similar to the above EPR measurement. In the
inflationary model, a scalar field, called inflaton, is introduced to yield
exponential cosmic expansion: $a\left( t\right) \propto e^{Ht}$, with $H$ is
almost constant. In this space-time, the scalar field fundamental mode $\nu
_{k}\left( t\right) $ obeys $v_{k}^{\prime \prime }+\left( {%
k^{2}-2H^{2}e^{2Ht}}\right) v_{k}=0$. This solves as $v_{k}\approx
k^{-1/2}\left( {1+ik^{-1}He^{Ht}}\right) $, which represents a strongly
squeezed state.

In the standard cosmological scenario, one simply adopts the identification
of the quantum correlation $\left. {\left\langle {\hat{\phi}\left( x\right) 
\hat{\phi}\left( y\right) }\right\rangle }\right\vert _{k}$ as statistical
fluctuation $\left\vert {\delta \varphi \left( k\right) }\right\vert ^{2}$
assuming that the fluctuation modes become classical when they cross the
horizon. Then, the power spectrum is given by $P_{\delta \varphi }\approx
\left( H/\left( 2\pi \right) \right) ^{2}$ at the horizon crossing time.
This is a scale invariant spectrum which simply reflects the fact that all
the modes evolve analogously in de Sitter space.

However actually, the above squeezed state is in the quantum state which
still respects the spatial translational invariance. If there were no
physical measuring process or a spontaneous symmetry breaking process, never
arises the classical fluctuation pattern which violates the symmetry. The
most promising process for this role would be the nonlinear interaction at
the reheating era just after the inflation. In the simple $\lambda \phi ^{4}$
model, the k-mode mean field $\varphi _{k}$ acts as an order parameter of
the measuring apparatus.

The generalized effective action

\begin{equation}
\tilde{\Gamma}=\tilde{\Gamma}_{0}+\int \lambda \tilde{\varphi}_{0}^{2}\delta 
\tilde{\varphi}\left( k\right) \left\langle {\tilde{\phi}\left( x\right) 
\tilde{\phi}\left( y\right) }\right\rangle \lambda \tilde{\varphi}_{0}^{2}%
\tilde{\varphi}\left( {-k}\right) dk+...
\end{equation}%
has the real kernel 
\begin{equation}
\func{Re}\left\langle {\phi \left( \vec{z}\right) \phi \left( 0\right) }%
\right\rangle =\int_{0}^{\infty }{dk\frac{\sin [kz]}{4\pi ^{2}z}\left[ {%
\left( {\frac{{\eta }^{\prime 3}-\eta ^{3}}{3{\eta }^{\prime }\eta }}\right)
+O\left( {k^{3}}\right) }\right] \equiv }\text{ }{G}_{\text{ret}},
\end{equation}%
and the imaginary kernel 
\begin{equation}
\func{Im}\left\langle {\phi \left( \vec{z}\right) \phi \left( 0\right) }%
\right\rangle =\int_{0}^{\infty }{dk\frac{\sin [kz]}{4\pi ^{2}z}\left[ {%
\begin{array}{l}
\frac{-1}{{\eta }^{\prime 2}}-\frac{1}{2}\left( {\frac{\eta }{{\eta }%
^{\prime }}+\frac{{\eta }^{\prime }}{\eta }}\right) \\ 
+O\left( {k^{2}}\right)%
\end{array}%
}\right] \equiv }\text{ }G_{\text{c}},
\end{equation}%
where $\eta \equiv -1/\left( aH\right) $. The real kernel is infrared finite
and yields finite renormalization contribution disregarding the ultraviolet
manipulations. On the other hand, the imaginary kernel that yields
statistical correlation is infrared divergent. However this does not violate
the physical evolution of the order parameter since the stochastic force is
always finite at each moment. Contrary to the laboratory case, we did not
consider the system state because the measurement process now is an
absorptive type and the system quanta is lost after the interaction. The
relevant information in the system is, however, inherited by the order
parameter $\varphi \left( x\right) $.

Applying the least action principle on $\func{Re}\tilde{\Gamma}+\xi \varphi
_{\Delta }\equiv S_{\text{eff}},$ we have 
\begin{equation}
\ddot{\varphi}\left( x\right) +{V}^{\prime }\left( {\varphi \left( x\right) }%
\right) +\int_{-\infty }^{t_{x}}{dyG_{\mbox{ret}}\left( {x-y}\right) \varphi
\left( y\right) }=\xi \left( x\right) ,
\end{equation}%
where the real part of the effective action yields the infrared-finite
renormalization. The stochastic force $\xi (x)$ has the correlation$%
\left\langle {\xi \left( x\right) \xi \left( y\right) }\right\rangle =G_{%
\mbox{c}}\left( {x-y}\right) \lambda ^{2}\varphi _{0}\left( x\right)
^{2}\varphi _{0}\left( y\right) ^{2}$ derived from the imaginary part of the
generalized effective action. Since the most relevant part of the above
equation becomes $\ddot{\varphi}_{k}\approx \xi _{k}$, we have 
\begin{equation}
P_{\delta \varphi }=\frac{4\pi k^{3}}{\left( {2\pi }\right) ^{3}}\left\vert {%
\varphi _{k}}\right\vert ^{2}\mathrel{\mathop{\kern0pt
\longrightarrow}\limits _{k/\left({aH}\right)=1}}\lambda ^{2}\left( {\Delta
t\;\varphi _{0}}\right) ^{4}\left( {\frac{H}{2\pi }}\right) ^{2},
\end{equation}%
where $\Delta t$ is the time scale of the reheating era and $\varphi _{0}$
is the typical amplitude of the order parameter in this era.

In the present $\lambda \phi ^{4}$ case at the reheating era, $\Delta
t\approx \varphi _{0}/\dot{\varphi}_{0}\mbox{, and }V\left( {\varphi _{0}}%
\right) \approx 0$ yield 
\begin{equation}
P_{\delta \varphi }=O\left( 1\right) \left( {\frac{H}{2\pi }}\right) ^{2},
\end{equation}%
which is almost the same as the standard result. In this special case, large
amplitude of the order parameter $\varphi _{0}$ exactly cancels the small
time scale $\Delta t$ to yield the factor of order one in front of the scale
free power spectrum. On the other hand for example in the chaotic
inflationary model, small field $\varphi _{0}\approx 0$ at reheating may
yield very small fluctuations. Although the scale free property of the power
spectrum is quite robust, the amplitude generally depends on the interaction
at reheating.

\section{Conclusions}

Quantum measurement process is physical. Actually we constructed a simple
measurement model in which all the time evolution including the measurement
process is autonomously described by the set of Eqs.(\ref{phidot}, \ref%
{eqrho}). This model is based on the \textsl{spontaneous symmetry breaking
(SSB)} in the apparatus. On the other hand in the system, initially \textsl{%
entangled }system with the apparatus eventually \textsl{decohere }and then 
\textsl{procohere }with firm correlation between the system and the order
parameter.

This method is applied to EPR measurement of spins in laboratories and the
effective action method correctly yields the correlation of the two
detectors. This correlation is fully consistent with quantum mechanics and
the violation of Bell inequality is observed. The essence of the process
here is the SSB among the two equivalent detectors, as well as SSB within
each detector.

On the other hand in the early Universe, the very similar process could
spontaneously break the spatial translational symmetry to yield density
fluctuations as classical statistical object. In this case, the SSB process
at the reheating when the nonlinearity enters is essential and this
nonlinearity determines the strength of the density fluctuations while the
scaling in the power spectrum is unchanged.

This symmetry breaking of the spatial translation is indispensable for the
generation of the classical density fluctuations in the early Universe.
Popular quantum decoherence process itself does not break this symmetry. The
quantum mode which goes outside the horizon does not break this symmetry. In
the present paper, the SSB of this symmetry was successfully described by
the generalized effective action method with its imaginary component due to
the squeezing of the state and the nonlinearity through the reheating
process.


\begin{thebibliography}{9}
\bibitem{BEC} T. Fukuyama and M. Morikawa, Prog. Theor. Phys. \textbf{115}
1047 (2006); Phys.Rev.\textbf{D80}:063520 (2009).

\bibitem{pad1993} T. Padmanabhan, \emph{Structure Formation in the Universe}%
, Cambridge University Press (1993).

\bibitem{weinberg2008} S. Weinberg, \emph{Cosmology}, Oxford University
Press (2008).

\bibitem{morikawa1987} M. Morikawa, Prog. Theor. Phys. \textbf{77} 1163
(1987).

\bibitem{leon2011} G. Leon et al, arXiv:1107.3054v1 (2011).

\bibitem{morikawa1986} M. Morikawa, Phys. Rev. \textbf{D33} 3607 (1986).

\bibitem{morikawa1995} M. Morikawa, Prog. Theor. Phys. \textbf{93,} 685
(1995).

\bibitem{morikawa2006} M. Morikawa and A. Nakamichi, Prog. Theor. Phys.%
\textbf{\ 116}, 679 (2006).

\bibitem{einstein1935} A. Einstein, B. Podolsky, and N. Rosen, Phys. Rev. 
\textbf{47} 777 (1935).
\end{thebibliography}
\end{document}